\documentstyle[aps,amssymb,amsmath,graphicx,epsf]{revtex}

\begin{document}
%\draft

\title{Order statistics for $d$-dimensional diffusion processes}
\author{ S. B. Yuste\thanks{e-mail: santos@unex.es},
L. Acedo\thanks{e-mail: acedo@unex.es} }
\address{
Departamento de F\'{\i}sica, Universidad  de  Extremadura, E-06071 Badajoz,
Spain}

\author{Katja Lindenberg\thanks{e-mail: klindenberg@ucsd.edu}}
\address{Departament of Chemistry and Biochemistry and Institute for
Nonlinear Science, University of California San Diego, La Jolla,
California 92093-0340}
\date{\today}
\maketitle

\begin{abstract}
We present results for the ordered sequence of first passage times of
arrival of $N$ random walkers at a boundary in Euclidean spaces of $d$
dimensions.
\end{abstract}

\vskip 20pt


It is customary to assume that the important behaviors of a
group of {\em independent} events can all be characterized by studying the
behavior of one such event. For instance, one focuses on the mean time
that it takes a single random walker to arrive somewhere even when there
are many such random walkers in the system provided the walkers are
independent.  However, it is clear that there are situations in which
one might be interested, for example, in the mean time for the {\em first}
of the walkers to arrive somewhere or, more generally, in the ordered
sequence of a particular outcome.  The statistics of ordered outcomes
clearly depends on the number $N$ of events even when these are
independent.  For example, the mean time for the {\em first} of $N$
walkers to arrive somewhere must clearly decrease with increasing $N$.
The interest in so-called order statistics has grown with the
development of experimental techniques that make it possible to follow
single events on the microscopic scale.

One way to characterize the behavior of $N$ random walkers is to focus
on the order statistics for arrival at a boundary, and,
in particular, the mean and variance for the first arrival of the first,
second, third, etc. walker out of a set of $N$. 
Most of the early literature dealt with mean arrival times in one
dimension~\cite{lindenberg,wsl83,YL96}.
More recent literature has extended
these concepts to fractal lattices~\cite{Yprl,Ypre,DK} and has attempted
to include higher-dimensional Euclidean lattices as well~\cite{DK,YA00}.
Some of these recent efforts~\cite{YA00} have relied on a parallel
literature about another quantity of interest in these problems, namely,
the distinct number of sites visited by $N$
walkers~\cite{larralde}.  In this approach the
first arrival time statistics are obtained through a conjectured
relation between arrival times and distinct number of sites visited.
Yer another approach~\cite{DK} relies on scaling arguments that are said to
be independent of the underlying environment and can be applied in any
dimension to both ordered and disordered structures.  However, the
results obtained in this work do not agree with any of the other
published results (nor with those obtained herein), even in the
well-established one-dimensional problem.  This difference
is apparent in the leading term of the result and also in the
form of the series that follows the leading term.  

In addition to order statistics and distinct number of
sites visited, the behavior of $N$ random walkers has also been
characterized in terms of the maximal excursion~\cite{bidaux,various}.  

In order to complete the $N$-independent-walker panorama, in this report
we present results for the ordered sequence of first passage times of
arrival at a boundary in Euclidean spaces of $d$ dimensions.   This
completion is made possible by a result reported in the maximal
excursion literature~\cite{bidaux}.  The leading terms
obtained here agree with those obtained earlier~\cite{YA00} and therefore
confirm the conjecture made in that work (at least to leading order).

The calculation of ordered sequences of first passage time moments 
is rather elaborate but has been laid out in detail in a number of
previous papers~\cite{wsl83,YL96,Yprl,Ypre}.  These steps are easy
to state but complex to carry out.  The $m$th moment of the $j$th
passage time (that is, of the first arrival of the $j$th particle out of
$N$) at a distance $r$ from the origin is
\begin{equation}
\langle t^m_{j,N}\rangle=\int_0^\infty t^m \psi_{j,N}(r,t) dt .
\label{deftmjN}
\end{equation}
Here $\psi_{j,N}(r,t)$ is the probability density of the time it takes
the $j$th particle to first reach the given distance $r$.  This quantity
can in turn be expressed in terms of the first-passage time density to
this distance of a single particle,
$\psi(r,t)\equiv\psi_{1,1}(r,t)$~\cite{wsl83,feller}:
\begin{equation}
\label{psijNt}
\psi_{j,N}(r,t)= N!/[(N-j)!(j-1)!]
\psi(r,t) h^{j-1}(r,t)  [1-h(r,t)]^{N-j} \;. 
\end{equation}
Here $h(r,t)=\int_0^t \psi(r,\tau) d\tau$ is the mortality function, i. e.,
the probability that a single diffusing
particle has reached the distance $r$ during the time interval $(0,t)$.
Thus, knowledge of the function $\psi(r,t)$ (or $h(r,t)$) in principle
allows the full evaluation of first passage time moments -- 
provided one can carry out the integral in Eq.~(\ref{deftmjN}).
Alternatively, one can calculate the generating function of the moments,
\begin{equation}
{\cal U}_{N,m}(z)=
\sum_{j=1}^N \langle t^m_{j,N}\rangle z^{j-1} \; ,
\end{equation}
and from this obtain the moments via a Taylor series expansion in $z$.

Herein, of course, lies the difficulty of the problem: neither can the
integral in Eq.~(\ref{deftmjN}) be done easily, nor can the generating
function be calculated easily.  Indeed, the literature is based on
rather elaborate expansions and/or renormalization group procedures to
obtain the generating function, methods that rely on there being a
very large number of walkers ($N$ large) and on
the walker of interest being one of the first few ($j\ll N$) so that
short times dominate the moments.  The formal results are similar from one
system to another (Euclidean, fractal), the differences arising because
of the differences in the mortality function and, specifically, in
the short time behavior of this function.

The mortality function (and particularly its short time behavior)
for one-dimensional systems was calculated in the
early work on order statistics~\cite{wsl83,YL96}.  For certain
deterministic fractal geometries this function was considered in
more recent work~\cite{Yprl,Ypre}.  In all of these, the short-time
behavior of $h(r,t)$ has a ``universal" form.  It depends only on the
combination $t/r^2$ (which we denote simply as $t$, understanding that
it is the distance-scaled time) and is given by
\begin{equation}
\widetilde{h}(t) \approx A t^\mu e^{-\beta/t^\gamma}(1+h_1t^\delta)
\label{universal}
\end{equation}
where the tilde just stresses the fact that this is an asymptotic
result for $t\rightarrow 0$.
The constants $A$, $\mu$, $\beta$, $\gamma$, and $\delta$
vary from one system to another.   

Although mortality functions for Euclidean systems 
have long been known and used in the random walk
literature for a variety of problems~\cite{bidaux,SL80,Ho48},
its short-time behavior for systems of dimension $d>1$ has
not been incorporated into the
order statistics context.  Perhaps not surprisingly, it
turns out to fit the pattern Eq.~(\ref{universal}) 
and therefore the existing theories can directly be applied these
systems.  In particular, the time Laplace transform of the
mortality function $h(r,s)={\cal L}h(r,t)$
(indicated by the same symbol as the function but with the argument $t$
replaced with the Laplace variable $s$) is~\cite{bidaux}
\begin{equation}
\label{hrs}
s h(r,s)=\frac{2^{1-d/2}}{\Gamma(d/2)} 
\frac{(2dr^2s)^{\frac{d-2}{4}} }{I_{d/2-1}(\sqrt{2dr^2s})}
\end{equation}
where $I_\nu$ is a modified Bessel function of order $\nu$, and
\begin{equation}
\label{Iasin}
I_{d/2-1}(z)=\frac{e^{z}}{\sqrt{2\pi z}} 
\left[ 1- \frac{(d-1)(d-3)}{8 z} + {\cal O}\left(z^{-2}\right)\right]
\end{equation}
for $z \gg 1$.  In writing (\ref{hrs}) an implicit choice of the
diffusion coefficient $D$ in the standard mean squared displacement
relation $\left< r^2\right> =2dDt$ has been made, namely, $D=1/2d$, so
that $\left< r^2\right> =t$ in each Euclidean dimension.
The inverse Laplace transform of
$s^\nu \exp(-a s^{1/2})$ is~\cite{roberts}
\begin{equation}
\label{Lainv}
{\cal L}^{-1} \left\{s^\nu e^{-a s^{1/2}}\right\} =
\frac{1}{2^{\nu+1/2} \pi^{1/2} t^{\nu+1}}  
e^{-a^2/(8t)} 
D_{2\nu+1}\left(\frac{a}{\sqrt{2t}} \right)
\end{equation}
where $D_{n}(x)$ is a parabolic cylinder function
(or Whitaker's function). Asymptotic expansion of this function for $x
\gg |n|$~\cite{abramo} leads for
$r^2/t\gg 1$ to
\begin{equation}
\label{Srt}
h(r,t)=\frac{2}{\Gamma(d/2)}
\left(\frac{dr^2}{2t}\right)^{-1+d/2}
e^{-\frac{dr^2}{2t}} 
 \left[1+\frac{(d-3)}{2d}\frac{t}{r^2}+
			{\cal O}\left(\frac{t^2}{r^4}\right) 
	\right]
\end{equation}
or, in terms of the scaled time $t/r^2 \rightarrow t$, to the small $t$
result
\begin{equation}
\label{htilde}
\widetilde h(t)\equiv A t^\mu e^{-d/(2t)}(1+h_1 t)
\end{equation}
with
\begin{eqnarray}
A&=&\frac{2}{\Gamma(d/2)}
\left(\frac{d}{2}\right)^{-1+d/2}  \\
\mu &=&1-\frac{d}{2}  \\
h_1&=&\frac{d-3}{2d}.
\label{coefi}
\end{eqnarray}
This is precisely of the form (\ref{universal}).

Since the steps leading to the moment expansions are well documented in
the literature once the form (\ref{universal}) has been
established ~\cite{wsl83,YL96,Yprl,Ypre}, we
only present the results.  For the $m$-th moment of the first passage
time of the first of $N\gg 1$ particles we find (the distance $r$ does
not appear explicitly because all times are scaled)
\begin{eqnarray}
\langle t_{1,N}^m \rangle =&&
\left(\frac{d}{2\ln \lambda_0  N}\right)^m
 \left\{ 1+\frac{m}{\ln \lambda_0  N}
  \left(\mu\ln\ln\lambda_0  N-\gamma \right) \right. \nonumber\\
&& + \frac{m}{2\ln^2 \lambda_0  N} 
      \left[(1+m)(\frac{\pi^2}{6}+\gamma^2)+2\mu\gamma- h_1 d
	\right.	       \nonumber \\
    && \left. \left. 
    - 2\mu\left(\mu+(1+m)\gamma\right)\ln\ln\lambda_0  N+
		     \mu^2(1+m) \ln^2\ln\lambda_0  N \right] +
      {\cal O}\left(\frac{\ln^3\ln\lambda_0  N}{\ln^3\lambda_0  N}\right) \right\}
					    \; ,
\label{t1Nm}
\end{eqnarray}
where
\begin{equation}
\label{la0}
\lambda_0 =
\frac{2}{\Gamma(d/2)} .
\end{equation}
The $m$-th moment of the first passage time of the $j$-th particle with
$j\ll N$ is
\begin{equation}
\langle t_{j,N}^m  \rangle \approx 
\langle t_{1,N}^m  \rangle+
\frac{d^m m}{2^m\ln^{m+1} \lambda_0N}\;
 \sum_{n=1}^{j-1} \frac{\Delta_n(m)}{n} 
\label{tjNm}
\end{equation}
where
\begin{eqnarray}
 \Delta_n(m)=&&
1+
\frac{m+1}{\ln\lambda_0N}
\left[(-1)^n\frac{S_n(2)}{(n-1)!}+\mu \ln\ln(\lambda_0N)-\frac{\mu}{m+1}-
\gamma \right] 
 + {\cal O} \left( \frac{\ln^2\ln\lambda_0N}{\ln^{2}\lambda_0N}\right)
\label{Delta}
\end{eqnarray}
and $S_i(n)$ is a Stirling number of the second kind~\cite{abramo}.
In particular, the variance $\sigma_{j,N}^2 \equiv 
\langle t^2_{j,N}\rangle -\langle t_{j,N}\rangle^2$ 
can be obtained from Eq.~(\ref{tjNm}) with (\ref{t1Nm}):
\begin{eqnarray}
\sigma_{j,N}^2 =&&
\left(\frac{d }{2 \ln^{2} \lambda_0 N}\right)^2 
\left[ 
\frac{\pi^2}{6}-\left(\sum_{n=1}^{j-1}\frac{1}{n}\right)^2+
\sum_{n=1}^{j-1}(-1)^n \frac{2 S_n(2)}{n!} \right] \; 
\left[1+ {\cal O} \left( \frac{\ln^3\ln\lambda_0N}
{\ln \lambda_0N}\right)\right].
\label{sigmajN2}
\end{eqnarray}

To check on the range of validity of these results it would be
desirable to carry out direct computer simulations involving a
very large number of walkers, an exercise that is costly.  
Another approach is to integrate numerically the exact
Eq.~(\ref{deftmjN}) with Eq.~(\ref{psijNt}) and the exact mortality
function in the integrand for values of $t$ beyond the range of
validity of the short-time expansions.  The exact survival probability
in one dimension is~\cite{SL80}
\begin{equation}
\label{Srt1d}
S(r,t)=\frac{4}{\pi}
\sum_{n=0}^\infty \frac{(-1)^n}{2n+1} 
\exp \left[-
	\frac{(2n+1)^2 \pi^2 t}{8r^2}.
      \right]
\end{equation}
In two dimensions~\cite{SL80}
\begin{equation}
\label{Srt2d}
S(r,t)=
\sum_{n=0}^\infty \frac{2}{x_{0n} J_1(x_{0n})} 
\exp \left[ -
	\frac{x_{0n}^2 t}{4r^2}
      \right]
\end{equation}
where $J_1$ is the first Bessel function and $x_{0n}$ is
the $n$-th root of the Bessel function $J_0$.
For $d=3$ one obtains the Hollingsworth distribution~\cite{Ho48}
\begin{equation}
\label{Srt3d}
S(r,t)=1-\left(\frac{6 r^2}{\pi t} \right)^{1/2}
\sum_{n=-\infty}^\infty 
\exp \left[
	-\frac{3(2n+1)^2 r^2}{2 t}
      \right]
\end{equation}
To carry out the integration it is convenient to integrate by parts and
so that 
\begin{equation}
\langle t^m_{j,N}\rangle=\langle
t^m_{j,N}\rangle+ m \frac{N!}{(N-j)! j!}
\int_0^\infty t^{m-1} [1-S(r,t)]^j S(r,t)^{N-j} dt 
\label{deftmjNpp}
\end{equation}
where 
\begin{equation}
\langle t^m_{1,N}\rangle=
 m \int_0^\infty t^{m-1} S(r,t)^{N} dt .
\label{deftm1Npp}
\end{equation}
Note that these integrations become rather awkward for large values of $N$.

\begin{figure}[htb]
\begin{center}
\leavevmode
\epsfxsize = 2.3in
\epsffile{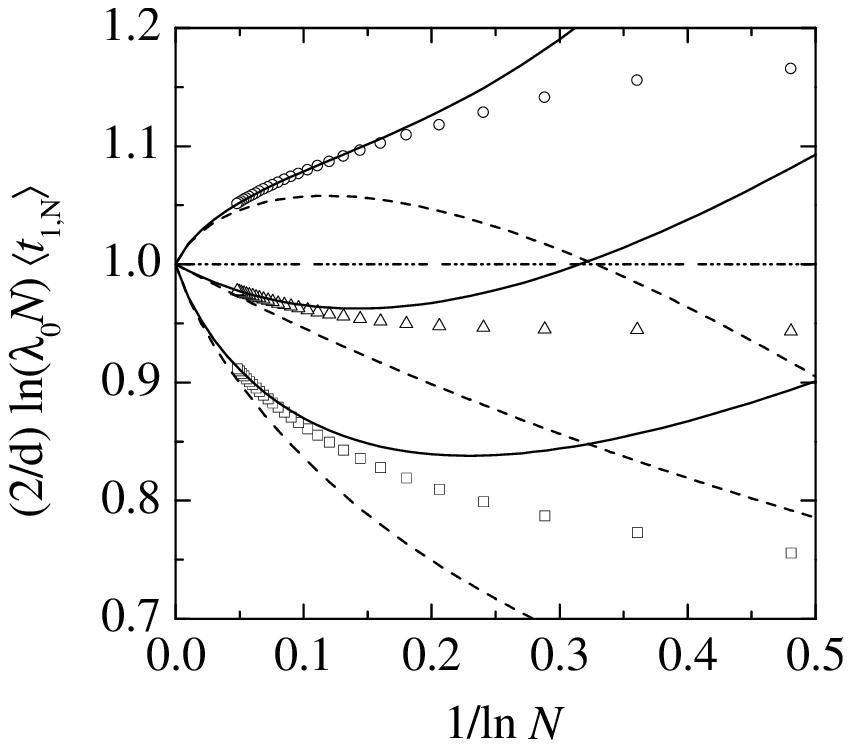}
\leavevmode
\epsfxsize = 2.3in
\epsffile{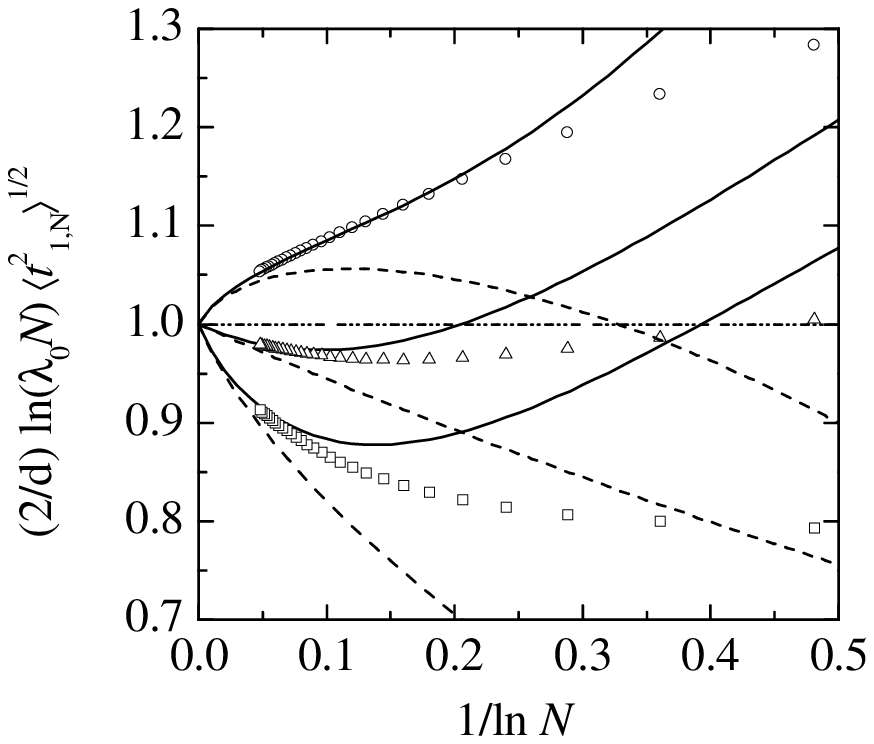}
\leavevmode
\epsfxsize = 2.3in
\epsffile{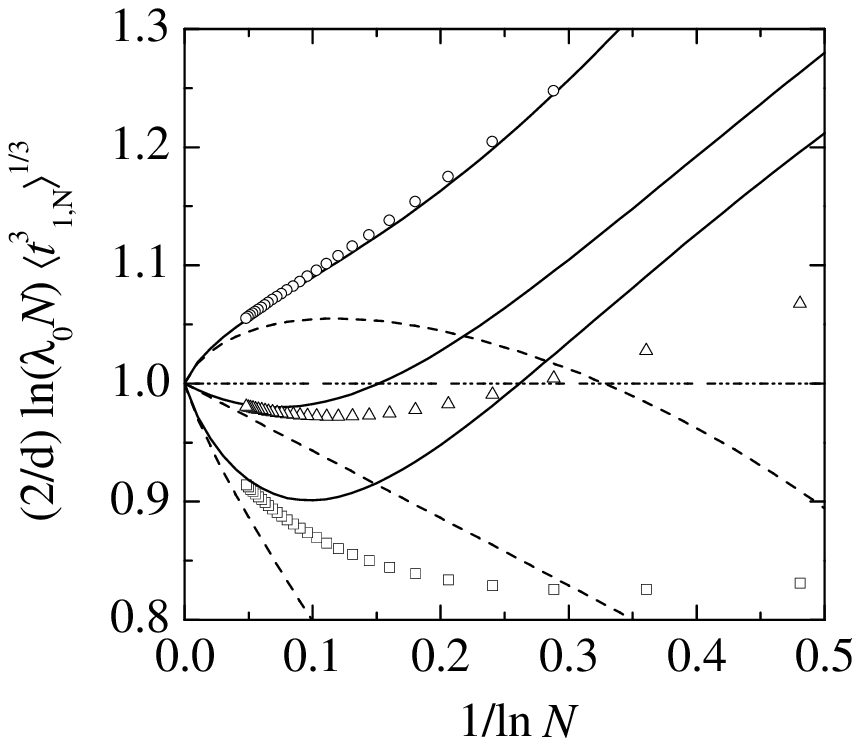}
\end{center}
\caption
{Scaled moments for first arrival of the first of $N$ particles at a
prescribed boundary as a function of $1/\ln N$.  First panel: first
moment (i. e., mean first passage time); second panel: second moment;
third panel: third moment.  Numerical results are indicated by circles
($d=1$), triangles ($d=2$), and squares ($d=3$).  Asymptotic results (cf.
Eq.~(\ref{t1Nm})) to zeroth order, first order, and second order are
shown by dotted curves, dashed curves, and solid curves respectively.
}
\label{t1Nmomentm}
\end{figure}

The numerical results and comparisons with asymptotic expressions
are presented graphically in a set of figures.  In Fig.~\ref{t1Nmomentm}
we present scaled results for  the $N$ dependence of
$\left< t_{1,N}^m\right>^{1/m}$, that is,
the $1/m$th power of the $m$th moment of the first passage time of the
first of $N$ particles.  The exact results obtained from the integration
(\ref{deftm1Npp}) are indicated by symbols and the theoretical results
of various orders by lines.  Circles denote exact results for $d=1$,
triangles for $d=2$, and squares for $d=3$.  The first panel shows
results for $m=1$, that is, for the mean first passage time of the first
particle to the desired boundary; the second panel presents the second
moment ($m=2$), and the third panel the third moment.  The moments on
the ordinates are scaled so that the scaled moment approaches 1 as
$1/\ln N \rightarrow 0$.  The integrations were performed for $N=2^3,
2^4,\cdots, 2^{30}$.  The dotted curves correspond to the asymptotic
results Eq.~(\ref{t1Nm}) to zeroth order, that in, only the first term in
the series.  The dashed lines include two terms, the solid lines three.
Clearly the convergence to the exact results improves with order
retained, but more slowly with increasing dimension.  In any case, the
deviations even at order 2 are clear on the scale of the figure at
around $N=100$.  

\begin{figure}[htb]
\begin{center}
\leavevmode
\epsfxsize = 3.0in
\epsffile{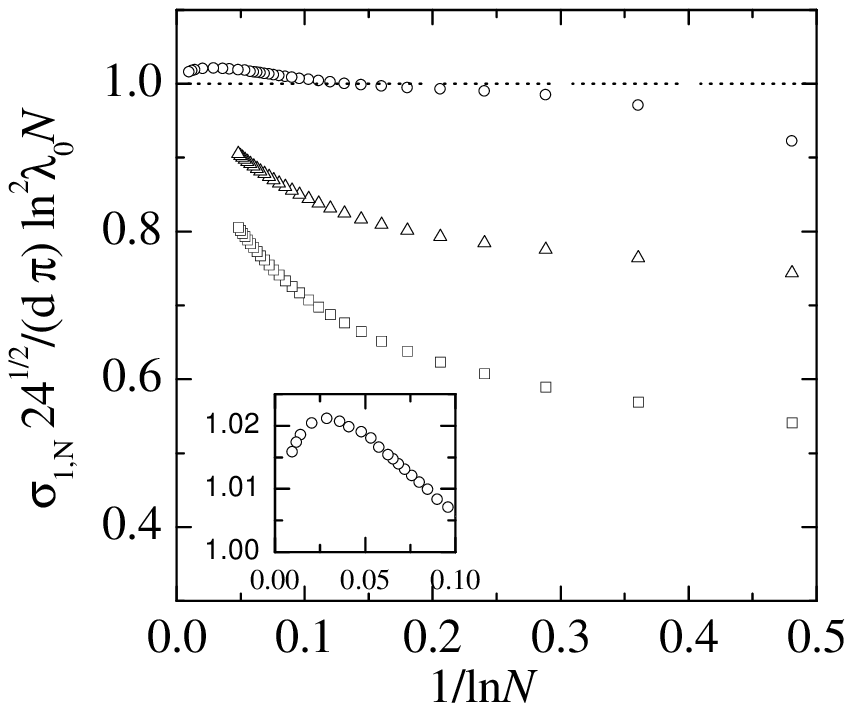}
\leavevmode
\epsfxsize = 3.0in
\epsffile{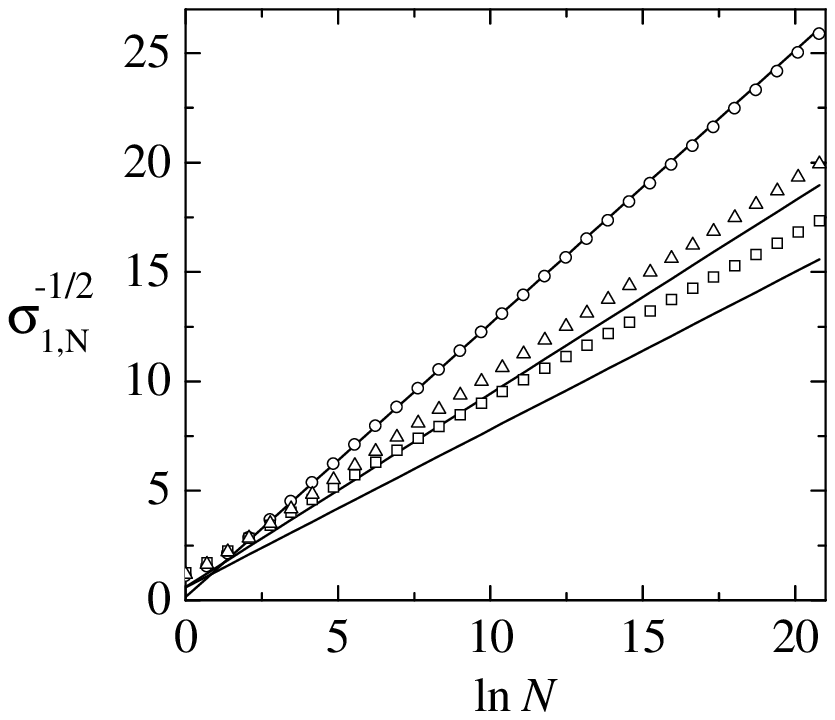}
\end{center}
\caption
{
$N$ dependence of the variance for first arrival of first of $N$
walkers at a prescribed boundary.  Symbols and lines are as in
Fig.~\ref{t1Nmomentm}.  First panel: variance scaled so that the zeroth
order asymptotic result is unity in all dimensions (indicated by the
dotted line).  Inset: detail of the $d=1$ numerical results.  Second
panel: same results, unscaled.
}
\label{variance}
\end{figure}

The two panels shown in Fig.~\ref{variance} deal with the variance
$\sigma_{1,N}^2 \equiv \langle t^2_{1,N}\rangle -\langle t_{1,N}\rangle^2$.
The leading term in the large-$N$ analysis of the variance
is given in Eq.~(\ref{sigmajN2}); for the variance we have only
calculated this leading (zeroth order) term.  In the first panel we
present the $1/\ln N$-dependent behavior of the variance scaled in
such a way that the
leading term in the theoretical expression gives unity for all the
dimensions $d=1, 2, 3$ (indicated by the dotted line in the figure).
The circles ($d=1$), triangles ($d=2$), and squares ($d=3$) are the
numerical results obtained from explicit integration.  A number of
interesting observations are apparent from this panel.  First, we see
that the leading term in the expansion of the variance leads to adequate
results for some range of $N$ only in one dimension.  In two and three
dimensions even for extremely large $N$ it is necessary to go beyond
zeroth order to obtain adequate analytic results.  It is thus clear that
the correction terms for the variance are much more important than for
the first passage time moments.  This can be seen from our asymptotic
expansions since the ratio of first to zeroth order terms goes as 
$\ln (\ln N)/\ln N$ for the moments but as $[\ln (\ln N)]^3/\ln N$
for the variance.  In fact, the second actually {\em grows} with
increasing $N$ at first, becoming larger than unity, and only begins
to decrease for extremely large values of $N$ (of order $2^{30}\sim 10^9$).  
This in turn leads to the ``anomaly" observed in the first panel and
enlarged in the inset, namely, that the exact variance
actually crosses and becomes {\em larger} than the zeroth order
theoretical one before settling down to the asymptotic value at
extremely large $N$.  (We only carried the numerical calculations to
these very large $N$ values for one-dimensional systems and in fact
reach the limit of numerical reliability in that region.)

The second panel in Fig.~\ref{variance} shows the same information as
the first but plotted in a different way to stress other features.  It
is simply a plot of the $\sigma_{1,N}^{-1/2}$ vs $\ln N$ comparing
numerical (symbols) and zeroth order asymptotic (lines) results in one,
two, and three dimensions.  Again, it is clear that for $d=1$ the
zeroth order asymptotic result is quite good but for higher dimensions
it is not adequate. 

In conclusion, we have filled some missing pieces in the mosaic of
results for the order statistics of $N$ independent random walkers in
$d$-dimensional lattices.  These results complement and confirm previous
conjectures, indicate that large-$N$ asymptotic expansions converge more
rapidly in lower dimensions, and that convergence of higher cumulants
such as variances is more problematic than that of first passage time
moments.

%\section{Acknowledgments}
This work has been supported in part by the Ministerio de Ciencia y
Tecnolog\'{\i}a (Spain) through Grant No. BFM2001-0718
and by the Engineering Research Program of
the Office of Basic Energy Sciences at the U. S. Department of Energy
under Grant No. DE-FG03-86ER13606.  SBY is also grateful to the DGES
(Spain) for a sabbatical grant (No.\ PR2000-0116) and to the Department
of Chemistry and Biochemistry of the University of California, San Diego
for its hospitality.

\end{document}